\documentclass[a4paper, 12pt]{article}
\usepackage{amsfonts}
\usepackage{amsbsy}
\usepackage{amsmath}
\usepackage{amsthm}

\newtheorem*{theorem}{Theorem}

\newcommand{\p}{\mathbb P}
\newcommand{\C}{\mathcal C}
\newcommand{\Ce}{\mathcal C_e}
\newcommand{\K}{\mathcal K}
\newcommand{\Hi}{\mathcal H}

\newcommand{\ket}[1]{|#1\rangle}
\newcommand{\bra}[1]{\langle#1|}

\begin{document}

\title{\bf{Partial Observables in Extended Systems}}

\author{\large
Frank Hellmann\footnote{pmxfh@nottingham.ac.uk}
 \\[3mm]
\em
\normalsize
{University of Nottingham}}

\date{\small\today}
\maketitle

\begin{abstract}

\noindent We consider ``unphysical'', kinematic observables that do not commute with the constraints of a gauge system in the context of an extension of the system. We show that these observables, while not predictable, can nevertheless be said to have a physical interpretation. They implement Rovelli's concept of partial and relational observables. We investigate the propositional structure of these observables and point out interpretational issues. We find that to make relational statements in the quantum theory one must deal directly with these observables. In particular we argue that in this scenario the spectra of kinematic observables are what is experimentally accessible.

\end{abstract}

\section{Introduction}

It is generally accepted that in a theory with gauge symmetries only quantities which commute with the constraints that generate the gauge transformations can be predicted unambiguously by the theory \cite{DiracBook, ThiemannBook}. It is a distinct, but often implicitly inferred statement that thus these are the only quantities that have a physical interpretation. Indeed the degrees of freedom that don't commute with the constraints are usually called unphysical. The key observation of Rovelli in \cite{Rovelli:2001bz} was that at least in some situations kinematic quantities that don't commute with the constraints have a physical interpretation as ``apparatus readings''. In general it takes several of these readings taken together to make up a complete measurement of the state of the system, e.g. we need to read of the clock of the wall and apply the yardstick to determine that a particle is at location $x$ at time $t$. Only this joint information can be unambiguously predicted from the system itself, and therefore only the concurrent (understood in a logical, not temporal sense) determination of these apparatus readings should correspond to a Dirac observable. These kinematical observables interpreted as apparatus readings are called the partial observables of the system. However, the theory as presented in the literature gives no indication as to when this physical interpretation is viable. Furthermore severe ambiguities arise in the quantum mechanical case pertaining to the question what the physical spectra are, see e.g. \cite{Dittrich:2007th, Rovelli:2007ep}.

In the context of General Relativity the question of the viability of such a physical interpretation of kinematic quantities becomes particularly important as diffeomorphisms appear as gauge symmetries in the canonical formulation. As any local observable is not invariant under diffeomorphisms  this would imply that it is unphysical. In order to understand the localized measurements we perform to actually test General Relativity it is therefore definitely necessary to go, in some way, beyond the gauge invariant quantities of pure General Relativity, e.g.\cite{Giddings:2005id, Rovelli:1990ph} and references therein\footnote{While relational localization using the gravitational degrees of freedom is possible, these are not the quantities that we actually measure to test General Relativity.}.

In this paper we will propose a definition of partial observables as certain Dirac Observables of extended systems. We will show that this identification allows us to resolve the above mentioned ambiguity. To make this paper self contained we will begin by reviewing the classical construction of partial and complete observables in section 2. In section 3 we will outline the quantization of these systems using group averaging techniques. We will give a new simple formal proof that this formalism is indeed a straightforward extension of the usual formalism of Quantum Mechanics. We will then show the ambiguity regarding physical spectra and propositions in the formalism in detail, and point out further open issues. In section 4 we give the proposed definition of partial observables as well as an argument that resolves the ambiguity. We conclude with a brief discussion on how these considerations apply to General Relativity.

\section{The Classical Theory}

We begin with a configuration space space $\C$ and its tangent bundle $T^*\C$, the phase space. We denote position and momenta by $q_i$ and $p_i$ respectively. The phase space then comes equipped with the standard Poisson structure: $\{q_i,q_j\} = 0$ and $\{q_i,p_j\} = \delta_{ij}$. As we will study the formal properties of the constraint structure we will, for clarity, look at fully constrained systems with a weakly vanishing Hamiltonian $H(q_i,p_i) = \lambda_j C^j(q_i,p_i) = 0$. This formalism is completely equivalent to the usual Hamiltonian time evolution formalism \cite{Rovelli:2004tv, Rovelli:2002ef}. To further simplify the discussion we will specify to systems with just one constraint, such that $H(q_i,p_i) = \lambda C(q_i,p_i) = 0$. Weak Dirac observables $O$ are those phase space functions which weakly commute with the constraint, that is $\{O,C\} = f C$, where $f$ is a smooth function on the phase space. Strong Dirac observables have $\{O,C\} = 0$ on the whole phase space. The flow under the vector field associated to the constraint through the symplectic structure is defined as $\frac{d}{d\tau} \alpha^{\tau}(\cdot)|_{\tau=0} = \{C,\cdot\}$. It traces out gauge orbits in phase space. The projections of these orbits onto the configuration space are then referred to as the motions of the system.\cite{Rovelli:2001bz}

Consider as an example the free particle of unit mass. The configuration space is $\mathbb{R}^2$ parametrized by $t$ and $x$, the constraint is given by $C = p_t + \frac{1}{2} p_x^2$. On the constraint surface the variable conjugate to $t$ is thus minus the particle energy. Neither $t$ nor $x$ commute with the constraint, they are not Dirac observables. This has a clear physical interpretation here, both of them are not determined independently. They depend on a further parameter, namely, when we measure the particle, which clearly is not predictable by the theory. Yet, for known momentum $p_x$ which is gauge invariant, a pair of values for both of these observables determined concurrently does fix the state of the system, which in turn can be predicted. Therefore, as stated in the introduction, we call $x$ and $t$ partial observables. We can then write their joint determination as a Dirac observable in the sense that given a state of the system, and given a particular value for the one, the other is predicted by the theory\footnote{Clearly this is the case only if the particle is not at rest, which we shall assume here.}. This Dirac observable can be defined rigorously also for the case of systems with infinitely many constraints \cite{Dittrich:2005kc, Dittrich:2004cb}. We shall use the integral representation (which was discussed by Marolf in \cite{Marolf:1994wh}) of the Dirac observable. For a general system the Dirac observable that gives $q_i$ if $q_j = \beta$ is given by:

\begin{equation}
	q_i|_{q_j = \beta} = \int{\alpha^{\tau}(q_i) \{C,\alpha^{\tau}(q_j)\} \delta(\alpha^{\tau}(q_j) - \beta) d\tau}
\end{equation}

Here $d\tau$ is the translation invariant measure on the gauge group. For our example of the free particle we have:

\begin{eqnarray}
x|_{t = \beta} = x + p_x (\beta - t) &&\nonumber\\
t|_{x = \beta} = t + \frac{1}{p_x}(\beta - x)&&
\end{eqnarray}

These clearly commute with the constraint. For the purpose of classical mechanics it is therefore entirely equivalent to consider either the complete observables just given, or the pairs of values of the partial observables as measurements of the system in question. In both cases there is a dependence on an element outside of the system itself. In the case of the Dirac observables this is the external parameter $\beta$. Due to this parameter this kind of observables was originally called an Evolving Constant of Motion. In the case of the partial observables this dependence is in what particular pair of values along the motion of the system we obtain. Whether to consider the Dirac observable or the partial observable is merely a matter of convenience, both give the same predictions for the theory.

If the surface defined by $q_j = \beta$ is only intersected by the motions once this discussion is complete. If multiple intersections are possible we have to deal with multivalued complete observables in general. For the specific representation of complete observables given here this is resolved by summing over all intersections.

\section{The Quantum Theory}

\subsection{Group Averaging and ordinary Quantum Mechanics}

To quantize these kind of systems we shall use the language of the group averaging implementation of the Dirac prescription \cite{Marolf:2000iq, Giulini:1999kc}, that is, we start with a kinematical Hilbert space $\K = L^2(\C)$ with the standard representation of our phase space variables as operators. This immediately gives us access to the kinematical operators corresponding to partial observables. Therefore this quantization scheme is well suited to illuminate the issues we want to discuss. The physical Hilbert space $\Hi$ is then defined ala Dirac by the condition that $\hat{C}\ket{\psi} = 0$. On $\K$ we define what is often called the ``projector'' onto the physical Hilbert space by:

\begin{equation}
	\p = \int{d\tau \exp(i \tau \hat{C})} = \int{d\tau U(\tau)}
\end{equation}

Where $d\tau$ is the same translation invariant measure on the gauge group used before. This will only be a projector if the gauge orbits induced by $C$ are compact, but the complications induced by noncompactness do not bear upon our discussion here. Formally it is immediate that $\p$ projects onto the physical Hilbert space as:

\begin{eqnarray}
  &\exp(i \beta \hat{C}) \ket{\psi} = \ket{\psi} \; \Leftrightarrow \; \hat{C} \ket{\psi} = 0&\nonumber\\
  &\exp(i \beta \hat{C}) \p \ket{\psi} = \int{d\tau \exp(i (\tau + \beta) \hat{C})} = \p \ket{\psi}&\nonumber\\
  &\Rightarrow \hat{C} \p \ket{\psi} = 0 \; \forall \psi&
\end{eqnarray}

We can then, for example, interpret the probability for a particular pair of values of partial observables as arising from a POVM (Positive Operator Valued Measure) constructed from operators of the form $\p P^{q_i}_\alpha P^{q_j}_\beta \p$. Here $P^{q_j}_\beta$ is the projector on the eigenspace of $q_j$ associated with the eigenvalue $\beta$. For more rigorous definitions of this quantization scheme see the cited literature. Some headway toward giving a consistent probability interpretation for this kind of quantities by providing a version of the extension theorem known from Quantum Information Theory for the measurement of POVMs \cite{NielsenQCQI} was made in \cite{Hellmann:2006fd}, though the precise physics of this extension remain unclear. Up to operator ordering we can also give an immediate quantization of the complete observables described in equation (1) as:

\begin{eqnarray}
	\hat{q}_i|_{q_j = \beta} &=& \int{U(\tau) \hat{q}_i [C,\hat{q}_j] P^{q_j}_\beta U^\dagger(\tau) d\tau}\nonumber\\
\end{eqnarray} 

Thus again we have two pictures emerging, we can either evaluate pairs of kinematical operators or a Dirac observable constructed from two of them in which the value taken by one of the operators is implicit in an external parameter. In contrast to the POVMs mentioned above here we have a Hermitian operator on the physical Hilbert space and thus the standard rules of quantum mechanics can be applied.

The example of the free particle is now again straightforward to implement. The solutions of $\hat{C} \psi(t,x) = 0$ are precisely the solutions of the ordinary Schr\"odinger equation: $(i \partial_t - \hbar \frac{1}{2} \partial_x^2) \psi(t,x) = 0$. For a state in $\Hi$ that solves this equation the expectation value of the operator $\hat{x}|_{t=\beta}$ is then just:

\begin{equation}
	\bra{\psi}\hat{x}|_{t=\beta}\ket{\psi} = \bra{\psi}\hat{x}P^t_\beta \ket{\psi} \int{d\tau} = \int{dx \psi(\beta,x) x \overline{\psi(\beta,x)}} \int{d\tau}
\end{equation}

Which up to an infinite constant arising because we neglected the fact that the gauge orbits can be noncompact, is precisely the expectation value of ordinary quantum mechanics. For this correspondence to work it is crucial that we have localized the system using a good time variable, that is, using a kinematic operator such that $[\hat{t},\hat{C}] = 1$. Given such an operator we can quite generally show that all the ordinary predictions of quantum mechanics are captured by this formalism. To see this consider the operator $P^t_\beta \p P^t_\alpha$ if $[\hat{t},\hat{C}] = 1$ then clearly $U(\tau)tU^\dagger(\tau) = t + \tau$ this implies then that:

\begin{equation}
	P^t_\beta \p P^t_\alpha = \int{d\tau P^t_\beta U(\tau) P^t_\alpha U^\dagger(\tau) U(\tau)} = \int{d\tau P^t_\beta P^t_{\alpha + \tau} U(\tau)} = P^t_\beta U(\beta - \alpha)
\end{equation}

In fact by considering the matrix elements of the projector between kinematical states that diagonalize $\hat{t}$ it is straightforward to confirm that for a constraint of the form $C=p_t+H$ this unitary transformation is just the usual time evolution acting between the eigenspaces of $t$. We use the fact that $C$ operating on the image of $\p$ is zero:

\begin{eqnarray}
0 & = & P_t^\beta C \p P_t^{\beta^\prime} \nonumber\\
  & = &\bra{n,\beta} C \p \ket{n^\prime,\beta^\prime} \nonumber\\
  & = &\bra{n,\beta} (p_t + H) \p \ket{n^\prime,\beta^\prime} \nonumber\\ 
  & = &\hbar \partial_\beta \bra{n,\beta} \p \ket{n^\prime,\beta^\prime} + n \bra{n,\beta} \p \ket{n^\prime,\beta^\prime} \delta_{nn^\prime}
\end{eqnarray}

Here $\ket{n,\beta}$ are the (generalized) eigenstates of $H$ and $t$ and the hermitian conjugate equation gives the evolution in $\beta^\prime$. Therefore the operator $P_t^\beta \p P_t^{\beta^\prime}$ satisfies the differential equation for the usual propagator.\footnote{Note also that this perspective immediately avoids the issue of the noncompactness of the gauge orbit.} In particular then $P^t_\beta \p P^t_\beta = P^t_\beta$, that is, this operator is the identity on the time eigenspace. If a basis of an eigenspace of $\hat{t}$ spans the whole physical Hilbert space if projected onto it using $\p$ and the individual basis vectors remain mutually linearly independent we can interpret $P^t_\beta \p$ as a map between the physical Hilbert space and the eigenspace associated to $\beta$, it is up to a, possibly infinite, constant unitary as $P^t_\beta \p (P^t_\beta \p)^\dagger = P^t_\beta \int d\tau$.

This implies that for this case the kinematical operators corresponding to the partial observables that commute with $t$ are mapped to unitarily equivalent operators on $\Hi$. For this case the formula for complete observables given above can be 
written as $q_i|_{t=\beta} =  \p P^t_\beta \hat{q}_i (\p P^t_\beta)^\dagger$.

\subsection{Propositional Structure and Spectra}

The spectral decomposition of the complete observable and the POVM constructed from the partial observables agree in the presence of a good time variable, and just as in the classical case which one to use is merely a matter of taste. There is again no ambiguity. Propositions in the kinematic Hilbert space translate to related propositions in $\Hi$, and into the propositions (that is, projectors) that the complete observable decomposes into. However, this will not be the case if we use arbitrary partial observables to define where the measurement is localized. This is to be expected, the propositional structure underlying an observable in QM is implemented through the addition of projectors. The operator corresponding to ``$A$ takes the values $a_1$ or $a_2$'' is $P^A_{a_1} + P^A_{a_2}$. We can usually do this because the two statements refer to disjoint regions in phase space. In the absence of constraints and under the usual time evolution that implies that they are mutually exclusive. That is, the one being true immediately implies the other one is false, distinct regions of the usual phase space correspond to distinct states of the system. Therefore the sum above gives a projector and a proposition again.\footnote{It is certainly possible to define a  composition of noncommuting propositions, however this does not have a clean interpretation even in ordinary quantum mechanics \cite{IshamBook}.} Furthermore, time changes in a uniform global way under the evolution. As we just showed this implies that the associated projectors for physical statements commute as long as they do not pertain both momenta and position and can be added to give a new projector.

For fully constrained systems and localization using general partial observables however a host of interpretational problems appear. Disjoint regions in the extended phase space can now of course be intersected by the same physical state. For example, the fact that a particle is at location $x$ at time $t$ does not preclude it from being at $x$ again at a later time. Thus disjoint regions do not correspond to exclusive propositions, and we lose the additivity of propositions referring to them. Classically this problem only shows up as multi valuedness in complete observables, as discussed above, or in the fact that certain observables diverge or become undefined at certain points of the phase space (e.g. a particle sitting motionless at the point of a particle detector). However, quantum mechanically the situation is greatly exacerbated. Consider the path integral representation of Quantum Mechanics. Even if we have classical single valuedness on a particular hypersurface defined by a certain value of a partial observable, the paths we sum over will not obey the equations of motion and can intersect the surface multiple times. We thus have interference between different locations on said surface. Unless motions are forbidden from ever turning around to come back to the said surface this will interfere with the orthogonality of projectors associated to distinct regions on the surface. The condition that motions can not come back through the surface in question is, however, precisely what defines a good time variable classically. Thus heuristically one would expect that the orthogonal propositions associated to the spectral decomposition of the complete observable would correspond to local orthogonal propositions if and only if the localization happens through a good time observable.

Indeed if the spectral decomposition of the complete observable is local, that is, if it can be understood through projectors into the subspace associated with the particular localization, this implies that the physical eigenstates of the complete observable can be unitarily mapped to kinematical states on the subspace in question. Then we can choose a basis in the rest of the kinematical Hilbert space whose elements are mapped by projection to the same physical eigenstates. This in turn enables us to write the physical projector as a sum of unitary evolution maps between complete commuting sets of these projectors. This does not quite imply that we have a good time variable in the system but it clearly is not the case in general. The precise obstructions toward a complete observable having a local propositional structure in the above sense remain unclear at this point.

This furthermore implies that the spectral decomposition of the complete observable discussed above has no obvious relationship to the POVM constructed from the pairs of kinematical operators. Consider that in the example of the particle elements of the POVM can refer to different instants of time, they will certainly not be orthogonal. Even if we construct projectors from the elements of the POVM these will then not mutually commute while the propositions arising in the spectral decomposition of the complete observable always must. While in the case of the POVM we are faced with the problem of having to give a consistent probability interpretation, the complete observable thus suffers from the deficiency that while it is supposed to have an interpretation as a kinematically localized statement, the propositions arising in its spectral decomposition, not being related to partial observables, seem to carry no such interpretation at all. If we decompose the Dirac observable in local quantities by using $\hat{q}_i|_{q_j = \beta} = \sum \gamma \, \hat{P}^{q_i}_\gamma|_{q_j = \beta}$ we face even graver interpretational issues as these operators now no longer need to be even positive definite. Even if we, for example, use a construction to absorb the negative parts of the spectra into the coefficients we would still be facing the issue of interpreting a POVM again.

Finally it was shown by Dittrich and Thiemann recently in \cite{Dittrich:2007th} that the spectrum of the complete observable can differ drastically from the spectrum of the partial observable used to construct it. Therefore beyond difficulties in the interpretation of the propositional structure of the theory we very concretely see an ambiguity in the spectrum the theory predicts for a certain measurement depending on whether we take the point of view of partial or complete observables as quantum mechanically meaningful in the absence of a good time variable.\cite{Rovelli:2007ep}

\section{External and Relational Degrees of Freedom}

\subsection{The Structure of Partial Observables in Extended Systems}

As we said in the introduction, the kinematical observables we are considering are not predictable by the theory of the system, but they do carry a physical interpretation. We have additional information that is implicit either in the external parameter of the complete observable or in the particular pair of kinematical quantities we get out of the set of pairs representing the same motion. This implies that at least in principle this information should be predictable by a more comprehensive theory. That is to say, in this larger system, the physical state should code not just information about what we measured but how we measured it as well.

Consider our system in question as a part of a larger system. We have a new phase space as $T^*\C \times T^*\Ce$ with the product Poisson structure. This phase space carries a single constraint, acting on both sets of variables jointly, which we assume to split into a sum as $C_{tot} = C + E$. Now we can identify three types of degrees of freedom, those of the system, living in $T^*\C$ and invariant under $C$, those external to the system living in $\Ce$ and invariant under $E$. Finally we observe that restricting ourselves to the two subsystems drops a degree of freedom. If $C$ is $n$ dimensional and $\Ce$ is $m$ dimensional, the subsystems each considered individually with $C$ and $E$ acting on them have $n-1$ and $m-1$ degrees of freedom. The joint system has $n+m-1$ degrees of freedom though. Thus one degree of freedom pertains to both systems jointly and can not be determined on their respective constraint surfaces.\footnote{The Dirac Observables of the subsystem used to measure the degrees of freedom defined here are of uniquely defined, as they can differ off the constraint surface of the subsystem terms of the form $f C$. There they can be sensitive to the additional degree of freedom. At least in the proximity of the constrained surface we can use $C$ and its conjugate variable to parametrize this dependence by Darboux's theorem and choose the Gauge equivalent version that does not depend on $C$ and its conjugate, and thus does not register the additional freedom of the joint system. This is what we mean precisely with the additional degree of freedom not being measurable on the constraint surface.} It is not invariant under the action of $C$ or $E$ but only under their joint action. This in some sense gives the relative position of the two systems to each other. It can change without affecting the gauge invariant degrees of freedom of each separate subsystem. This kind of relational degrees of freedom are then natural candidates for Dirac observables of the extended system corresponding to those kinematic quantities of the subsystem that have a physical interpretation.

Let us assume that we have a good time observable $e$ in $\Ce$, then we can use this to construct the Dirac observables $q_i|_{e=c}$, which we will abbreviate as $q_i^c = q_i|_{e=c}$. Clearly these are not invariant under either $C$ or $E$ but only under the joint action $C_{tot}$. Still, from this point on we are working exclusively with Dirac observables of the joint system. We can now consider again pairs of such observables, $q_i^c$.  Though these observables individually are not invariant under the constraint of the subsystem, they do give us information about the state of the subsystem when taken together. Consider again the simple example of the parametrized particle and take as environment an external clock. Then a joint observation of $t^c = t - (e - c)$ and $x^c = x - p_x (e - c)$ will give us two bits of information, the Newtonian time (or the time shown by a clock carried by the particle) when the (external) clock shows $c$ and the location of the particle when the clock shows $c$. Taken together these do clearly give us all the information we need. Individually they change if we shift the time origin though, thus they also know something about the relative state of the clock and the system.

We can now again separate out the content of the pair of observables that refers exclusively to the subsystem and the relational content. To do so we repeat the construction above for a complete observable only this time using the Dirac observables $q_i^c$. Note that under the flow of $C$ these just change as $\alpha^{\tau}(q_i^c) = q_i^{\tau+c}$ as they are invariant under the flow of $C_{tot} = C + E$ and thus transform under $C$ just as under $E$. The latter though merely induces a shift in the time variable $e$ which can be absorbed into the constant $c$. Therefore our complete observable appears as:

\begin{equation}
	q_i^c|_{q_j^c = \beta} = q_i|_{q_j = \beta} = \int{q_i^c \{C,q_j^c\} \delta(q_j^c - \beta) dc}
\end{equation}

The independence of $c$ precisely corresponds to the fact that complete observables are gauge invariant observables of the subsystem under investigation. We see thus that in the case where the system is part of a larger system which includes observables with a physical interpretation that behave as time in the extended system (at least approximately) the concept of partial and complete observables is completely implemented.\footnote{Note though that the complete observable will in general not be of the form described in the last footnote, and thus will still depend on the relational information though.} A good physical interpretation of the kinematic quantities of the subsystem under investigation can be given, even if the subsystem itself does not allow for a good time variable. The physical picture emerging for partial observables is now that we prepare the joint system such that $q_i^c$ takes on a certain value and then determine $q_j^c$.

Note that it is not necessary here to regard the extended system as fundamental. It plays the role of the clock on the wall. By an argument due to Pauli a perfect clock is quantum mechanically impossible, but all we require here is that the clock behave as such to good approximation in the time frame in which we perform our experiment. Then its degrees of freedom effectively provide us with a a clock variable $c$ that allows us to define the physically interesting quantity $t^c$, that is, how much physical time has passed in the system when our lab clock has ticked to $c$. The inner workings of the clock are not probed by our observables, only the general kinematic setup comes into play. As we showed above that for such $c$ the formalism here is equivalent to the standard formalism of Quantum Mechanics we might as well consider it a classical variable in this context.

\subsection{A Quantum Mechanical No-Go Theorem for Complete Observables}

Crucially this construction also allows us to shed some light on the quantum mechanical ambiguity on whether to use complete or partial observables in the construction of physical statements. A complete observable derives its interpretation from the localization that enters its construction. If the picture of partial observables that provide this localization is physical in an extended system, as in the case just examined, then this localization corresponds to physical gauge invariant information in the extended system. If the interpretation inherited is to be viable, the extended formalism needs to show us not just in what state we found the system, but also where we looked for it. Therefore besides $q_i|_{q_j = \beta}$ we need an observable that gives us the information ``$q^c_j = \beta$''. If ``$q^c_j \neq \beta$'' then the phase space function defined by the complete observable will have a different interpretation. The natural choice is of course $q_j^c$ itself. However there is no reason to assume that $\{q_i|_{q_j = \beta},q_j^c\}$ vanishes. In fact, in general it will not. Consider again the particle, while the complete observable $x|_{t=\beta}$ commutes with $t$, $t|_{x=\beta}$ does not at all commute with $x$ instead giving:

\begin{equation}
\{t|_{x=\beta},x\} \sim \frac{1}{p_x^2}
\end{equation}

Furthermore we can see that it is in general not possible to chose a different partial observable that gives the required interpretational information for the complete observable and commutes with the complete observable. Such a partial observable would have to satisfy $\{O_p, C\} = \{q_j^c, C\}$ to track the changes in the state of the relational degree of freedom in the same way and lead to the same functional form of the complete observable. Then $O_p - q_j^c = F$ is a Dirac observable of the subsystem depending only on its gauge invariant degrees of freedom. Then $\{q_i|_{q_j = \beta},O_p\} = \{q_i|_{q_j = \beta},q_j^c\} + \{q_i|_{q_j = \beta},F\}$. That is, its commutator differs from the commutator of the original partial observable by a gauge invariant observable again. Therefore unless $\{q_i|_{q_j = \beta},q_j^c\}$ is a complete observable there is no such $O_p$

This constitutes the main observation of this paper:

\begin{theorem}
Assume a system of the form $T^*\C \times T^*\Ce$ with a constraint that splits as $C_{tot} = C + E$, $C$ depending only on $T^*\C$ and $E$ only on $T^*\Ce$. For any complete observable $q_i|_{q_j = \beta}$ of the subsystem $\C$ for which $\{q_i|_{q_j = \beta},q_j\}$ is not a Dirac observable, no Dirac observable $O_p$ of the total system exists such that $\{O_p, C\} = \{q_j^c, C\}$ and $\{q_i|_{q_j = \beta},O_p\} = 0$. Thus there is no observable that tracks the relational information as $q_j$ does and that has vanishing Poisson bracket with $q_i|_{q_j = \beta}$.
\end{theorem}

While classically this Poisson bracket provide no obstruction to the measurement of the partial observable and the complete observable they of course do prevent just that when promoted to commutators quantum mechanically. This impossibility to simultaneously diagonalize the observable used for relational localization and the complete observable is precisely what leads to the problem of interpreting the spectral decomposition of complete observables discussed in the last section. This implies that if the physical interpretation of partial observables is applicable to give a relational interpretation to kinematical quantities we can not separate out the where we measure from what we measure. Yet if it is not applicable in the first place the interpretation of the complete observables becomes questionable anyhow. We conclude that the ambiguity in whether to consider the partial observables or the complete observables quantum mechanically is resolved toward the partial observables in the scenario given above.

To clarify the physical implications of this observation consider the very simple example of a free particle of unit mass in two dimensions hitting a screen. The system is a bit too simple to discuss the concepts developed above in full but the salient features are present. The total constraint is $C_{tot} = p_t + p_x^2 + p_y^2$. Say we locate the screen at $x = \beta$ and we want to know what the observable is for the $y$ position of the particle. The idea of complete observables would have us construct the Dirac observable:

\begin{equation}
y|_{x = \beta} = y + \frac{p_y}{p_x}(a-x)
\end{equation}

But now we have:

\begin{equation}
\{y + \frac{p_y}{p_x}(a-x),x\} = - \frac{p_y}{p_x^2}(a-x)
\end{equation}

which is not a Dirac observable of the whole system. This implies that this complete observable can not be determined concurrently with an observable that says that the particle hit the screen at $x = \beta$. Physically this is to be expected. What actually happens is that we determine the time at which the particle hits the screen and evaluate the distribution on $y$ at the projection on $x = \beta$ at that moment. If we want to discard the time of arrival we take a weighted sum over all such times, but in the measurement setup the time is available as a quantity that could be recorded. Thus the complete observable does not correspond to the way this measurement is described in ordinary quantum mechanics. The time of arrival thus plays an analogous role to the relational degrees of freedom discussed in the abstract setup above. The failure of the complete observable to take this aspect of the experiment into account leads to inequivalent predictions.

\section{Discussion and Conclusions}

The investigation presented here is of course far from exhaustive. We have given one particular scenario in which the ideas of partial observables appear naturally in the formalism of gauge systems, and there might well be others. Conversely it is also clear that the additional physics one gains by considering a gauge system in its environment is not completely captured by the ideas of partial observables, another example of such physics would be the Aharonov-Bohm effect. Both issues are left to future studies. Despite these caveats, it is clear though that the obstruction to separating out gauge invariant information in relationally localized measurements presented here will persist whenever we are able to lift the kinematic observable algebra to a Dirac observable algebra of an extended system. The argument does not depend on the way this lift is achieved here.

Remarkably, in \cite{Brown:1994py} Brown and Kuchar have given a form of dust matter for General Relativity with the right properties to apply the above reasoning. Giesel and Thiemann have recently used these to construct the quantum theory using AQG\cite{Giesel:2007wn}. Our view is that their construction gives an explicit form of the partial observables. The examples considered above also suggest that for this interpretation, it is sufficient to consider the dust used there as an approximate macroscopic object, that does not need explicit quantization as its influence is completely governed by its simple behavior under gauge transformations. In particular it seems that we need the dust only where we make the measurement. For a gravitational wave system for example we can treat most of spacetime around the radiating system as empty, and just consider a patch called, for example, ``earth'' as dust filled where then a bouncing laser between two reflective dust particles will detect an oscillation in distance between these particles even though they themselves have not moved with respect to their ambient space time. This then would imply that in the theory of Loop Quantum Gravity there is a natural class of measurements of the degrees of freedom of the theory for which the spectrum of the kinematical should be what is measured physically.

In another direction Meusburger has analyzed the physically accessible quantities for an observer in pure 2+1 Gravity in \cite{Meusburger:2008iz}. This system provides a good arena to analyze in how far the ideas presented here apply to typical observations in Gravity. Furthermore it does not introduce any problematic matter content. As opposed to Brown-Kuchar dust there is no deparametrisation of the theory there though, so the theorem shown here does not immediately apply.

To make contact with experiments in General Relativity it seems unavoidable to consider quantities localized with respect to non gravitational degrees of freedom. This was emphasized already early on in the life of the theory through the hole argument. We have shown an explicit construction along these lines even in very simple toy models. We found a clear obstruction to separating out gauge invariant degrees of freedom of the subsystem under investigation from those degrees of freedom used to localize our measurements in the quantum mechanical case. We could conclude that a consistent interpretation in terms of relationally localized quantities seems only possible by taking the kinematic structure of the theory into account, in it we see a minimal imprint of how the subsystem separates out from the larger systems in which it is embedded, or reversely speaking, how it couples to them. In other words, the kinematics seem to encode how we can couple to the system to extract physical predictions. This implies that the spectra of area and volume operators in Loop Quantum Gravity, even though not gauge invariant, should be considered physical predictions of the theory.

\section{Acknowledgements}

I would like to thank John Barrett, Madalin Guta, Jorma Louko, Eric Martinez-Pascual, Catherine Meusburger, Carlo Rovelli and Johannes Tambornino for discussions and comments on a draft of the paper.

\end{document}